# THE STABILITY OF THE ORBITS OF EARTH-MASS PLANETS IN AND NEAR THE HABITABLE ZONES OF KNOWN EXOPLANETARY SYSTEMS

**Barrie W Jones, David R Underwood, P Nick Sleep**

*The Open University, Walton Hall, Milton Keynes MK7 6AA, UK, b.w.jones@open.ac.uk*

**ABSTRACT**

We have shown that Earth-mass planets could survive in variously restricted regions of the habitable zones (HZs) of most of a sample of nine of the 93 main-sequence exoplanetary systems confirmed by May 2003. In a preliminary extrapolation of our results to the other systems, we estimate that roughly a third of the 93 systems might be able to have Earth-mass planets in stable, confined orbits somewhere in their HZs. Clearly, these systems should be high on the target list for exploration for terrestrial planets. We have reached this conclusion by launching putative Earth-mass planets in various orbits and following their fate with a mixed-variable symplectic integrator.

## 1. INTRODUCTION

It might be some years before we know for certain whether any exoplanetary systems have planets with masses the order of that of the Earth. We have therefore used a computer model to investigate a representative sample of the known exoplanetary systems, to see whether such planets could be present, and in particular whether they could remain confined to the habitable zones (HZs). If so then it is possible that life is present on any such planets.

The HZ is that range of distances from a star where water at the surface of an Earth-like planet would be in the liquid phase. We have used boundaries for the HZ originating with Kasting, Whitmire, & Reynolds [1]. The inner boundary is the maximum distance from the star where a runaway greenhouse effect would lead to the evaporation of all surface water, and the outer boundary is the maximum distance at which a cloud-free $CO_2$ atmosphere could maintain a surface temperature of 273K. Because of simplifications in the climate model of Kasting et al., these distances are conservative in that the HZ is likely to be wider. For zero-age main-sequence stars (ZAMS stars) the HZ lies closer to the star the later its spectral type, and as the star ages the boundaries move outwards. We have revised the boundaries of Kasting et al by using a more recent model of stellar evolution [2].

To test for confinement to the HZ, we launch putative Earth-mass planets into various orbits in the HZ and use a mixed-variable symplectic integrator to calculate the evolution of the orbit [3]. The integration is halted when an Earth-mass planet comes within three Hill radii ($3R_H$) of the giant planet. This is the distance at which a symplectic integrator becomes inaccurate, and is also the distance at which severe orbital perturbation of the Earth-mass planet occurs. Confinement requires that the integration lasts for the full pre-set time, *and* that the semimajor axis of the Earth-mass planet remains within the HZ. The pre-set time depends on the shortest orbital period in the system. The integrator requires a time-step shorter than about 5% of this shortest period. For systems like Rho Coronae Borealis the orbital period of the giant planet is so small that the pre-set time must be comparatively short to keep the CPU time per integration less than 100 or so hours. We used 100 Ma (100 million years) as the standard pre-set time for this system, and also for Gliese 876 and Upsilon Andromedae. For the other systems we used 1000 Ma as the standard.

## 2. THE EXOPLANETARY SYSTEMS STUDIED

We selected nine systems that between them represent a large proportion of the 93 main-sequence exoplanetary systems confirmed by May 2003.

Fig. 1 (see end of paper) shows the important characteristics of these systems, ordered by increasing period of the innermost planet. The ZAMS habitable zone HZ(0) is shown shaded, and the boundaries of HZ(now) by vertical dashed lines. Each giant planet is shown by a black disc labelled with the value of $m\sin(i_0)$ in Jupiter masses $m_J$, where $m$ is the mass of the giant and $i_0$ is the inclination of the planet's orbit with respect to the plane of the sky. All but Gliese 876 have been observed only by Doppler spectroscopy, which yields $m\sin(i_0)$ rather than $m$. The motion of Gliese 876 has also been detected astrometrically, giving a value $i_0 = 37°$. At each giant planet the solid line shows the total excursion $2\Delta r$ of the giant due to its eccentricity. The dashed line extends to $(3R_H + \Delta r)$ each side of the giant when it has its minimum mass ($R_H$ is proportional to $m^{1/3}$). Further information is given in the caption. The information in Fig. 1 comes largely from Schneider [4], including the references he gives.

For an integration we have to put in an actual mass $m$ for the giant planet. This is equivalent to setting $i_0$ to some value. For each system we used $i_0 = 90°$, which gives the minimum value for $m$, and other values, for example 42°, corresponding to 1.5 times the minimum.

## 3. RESULTS

We have obtained the following results for some of the nine systems that we believe apply generally.
- The effect of the mass of the terrestrial planets has been explored from $m_{EM}$ to $8m_{EM}$, where $m_{EM}$ is the mass of the Earth plus Moon. The mass has little effect, so we restrict ourselves here to describing results with a planet of mass $m_{EM}$ and labelled EM.
- The presence of a second EM affects the outcome not primarily through the direct gravitational interaction between these two planets, but through close encounters between them resulting from the effect of the giant planets on each of their orbits separately. We therefore focus on results with just one EM.
- The inclination of EM has been explored, up to 20°. There is little effect in most cases.
- All planets are launched with zero mean anomalies but with various longitudes of the periastra. The outcome can be sensitive to the differences $\Delta\varpi$ between these longitudes, in that some differences result in early close encounters (EM comes within $3R_H$ of the giant planet), whereas other differences result in no close encounters within the pre-set integration time. In such sensitive cases the inclination can affect the outcome.

Here is a brief summary of the results on each system.

*Upsilon Andromedae and Gliese 876*

Inspection of Fig. 1 shows that the whole of HZ(0) and HZ(now) are traversed by $(3R_H + \Delta r)$, and so there would seem to be no chance of avoiding an early close encounter between EM and a giant. This expectation is borne out by the integrations, with close encounters no later than about 10 Ma, and usually much sooner. Our work on Gl876 preceded the discovery of the inner, small giant (Fig. 1), but this only makes matters worse. Note that with $i_0 = 37°$ (from astrometry), the actual giant masses are 1.7 times the minima shown in Fig. 1.

*Rho Coronae Borealis*

The giant is well interior to HZ(0), and $(3R_H + \Delta r)$ at minimum giant mass is small (Fig. 1). It was therefore not a great surprise to us that, at minimum giant mass, confined orbits for EM are found right across HZ(0), even at the 1:6 mean-motion resonance near the inner boundary of HZ(0). HZ(now) is further out and so here too confinement is ubiquitous. We were more surprised to obtain similar confinement at eight times the minimum giant mass.

The age of Rho CrB is about 6000 Ma, so confinement for the standard pre-set integration time of 100 Ma could be a poor indicator of confinement for 6000 Ma. We therefore extended a few integrations for pre-set times 2-3 times longer than the standard. Full-term confinement was obtained in every case. Moreover, there is no sign of any long-term secular increase in the eccentricity of EM that has been seen (in other systems) when a close encounter subsequently occurred well into the integration. Therefore, we believe that confinement for 6000 Ma is common, even ubiquitous, in the habitable zones of Rho CrB.

*HD52265*

This is another case with the giant planet interior to HZ(0), though $(3R_H + \Delta r)$ is larger (Fig. 1). Confinement was obtained at minimum mass except when the initial semimajor axis of EM, $a_{EM}(0)$, was less than about 0.9 AU i.e. except near the inner boundary of HZ(0), and except near the 1:4 mean motion resonance at 1.235 AU. At 1.5 times the minimum mass, confinement was obtained beyond 1.0 AU (except at the 1:4 resonance).

*47 Ursae Majoris*

This system has an inner giant at the outer boundary of HZ(now), and a less massive giant further off with a present eccentricity $e < 0.1$. In Figure 1 the value is 0.1. This value was used at $t = 0$ in the integrations, and also a value of $10^{-5}$. As well as minimum mass, the giants were given 1.5 times the minimum, in accord with estimates of $i_0$ for this system [5]. At 1.5 minimum mass, with a few exceptions, the two giants were launched with $\Delta\varpi_{G1G2} = 0°$ and 120°, and $\Delta\varpi_{EMG1} = 120°$ and 180°. This taught us that $\Delta\varpi_{G1G2} = 120°$ was less likely to lead to confinement. To err on the side of caution in our claims for confinement, at minimum mass we ran configurations other than 120° only if there was no confinement at 120°.

For both masses there was no confinement beyond 1.25 AU. This is unsurprising given the reach of $(3R_H + \Delta r)$ in Fig. 1. At minimum giant planet mass, at $a_{EM}(0) < 1.25$ AU, we obtained confinement at both eccentricities and at all values of the $\Delta\varpi$ except near some of the mean-motion resonances. At 1.5 times the minimum mass, with $e(0) = 10^{-5}$ the outcome is similar, though at $a_{EM}(0) < 1.20$ AU. At 1.5 times the minimum mass and with $e(0) = 0.1$, confinement was obtained only at $a_{EM}(0) < 1.15$ AU and only for some $\Delta \varpi$.

47 UMa is about 7000 Ma old. The pre-set integration time of 1000 Ma is a modest fraction of this, though when close encounters occurred they did so within 500 Ma in all but a small proportion of cases, and in these cases there were clear upward trends in the eccentricity of EM. Therefore, confinement for 1000 Ma is very likely to mean confinement for 7000 Ma.

Our conclusions on 47 UMa are in accord with work done by others, though this other work is significantly less extensive [6] [7].

*HD196050*

In this system the giant is a modest distance beyond HZ(now), but has a rather large minimum mass of 3.0 $m_J$, and is in a rather eccentric orbit, with $e = 0.28$, so the whole of HZ(now) is traversed by $(3R_H + \Delta r)$ (Fig. 1). Confinement is obtained only within 1.0 AU of the star at minimum giant mass i.e. near the inner boundary, and nowhere within HZ(0) at 1.5 times this minimum.

*HD216435 (Tau$^1$ Gruis)*

Here too the giant is a modest distance beyond HZ(now), but has a minimum mass of only 1.23 $M_J$, and a less eccentric orbit, with $e = 0.14$. This star, though only about the age of the Sun, is more massive than the Sun (1.25 solar masses) and is thus near the end of its main sequence lifetime. Consequently, HZ(now) is substantially further from the star than HZ(0) (Fig. 1). Confinement of EM is not expected in HZ(now), and is not found, not even in the region around the outer boundary of HZ(now).

At minimum giant mass confinement is obtained within 1.7 AU of the star, except near the 1:3 mean motion resonance at 1.250 AU. The inner boundary of HZ(0) is at 1.13 AU. Therefore, there may have been an opportunity for life to evolve in the HZ, but it might not have survived to the present.

*HD72659*

With HD72659 we have a system more like the Solar System than the other eight systems we have studied, with the giant planet well beyond HZ(now) though not as safely distant as Jupiter is in our system. Also, the giant is at least 2.55 times the mass of Jupiter and is in a more eccentric orbit (0.18 versus 0.049). Nevertheless, even at 1.5 times the minimum giant mass there is confinement throughout the HZs, except at the 1:8 mean motion resonance near the inner boundary of HZ(0), and at the 1:3 resonance just outside HZ(now). At the minimum mass there is stability at the 1:8 resonance.

*Epsilon Eridani*

Epsilon Eridani suffers from having a giant with one of the highest eccentricities among the known exoplanets, 0.608. Fig. 1 shows that $(3R_H + \Delta r)$ at minimum giant mass extends inwards to nearly reach the inner boundary of HZ(0). Note that HZ(now) is barely further out. This is partly because Eps Eri is a K2V star and therefore evolves more slowly than the Sun, and partly because it is young, only 500-1000 Ma. We were uncertain whether we would find confined orbits, but we explored the inner part of the habitable zone and the region immediately interior to it, including all the mean motion resonances, which are closely spaced. At such resonances there was a strong tendency for $\Delta \varpi_{EMG} = 180°$ to give confinement, and for $\Delta \varpi_{EMG} = 0°$ to fail to do so, a well-known result.

At minimum giant mass, the EM orbits that remain confined for the 1000 Ma pre-set time, are restricted to $a_{EM}(0) < 0.59$ AU. Within this distance there is an intricate pattern of outcomes, depending on whether $\Delta \varpi_{EMG} = 180°$ or $0°$, and whether $a_{EM}(0)$ is at a mean-motion resonance. Only within 0.44 AU is there general confinement. We also explored 1.39 times the minimum mass. This is because Eps Eri has a circumstellar dust ring that is seen to be elliptical. Assuming it to be circular we get an estimate of $i_0 = 46°$, that corresponds to the 1.39 multiplier. At this greater mass there is again general confinement within 0.44 AU, but no confinement at all at $a_{EM}(0) > 0.46$ AU.

Further details of our work on Ups And, Gl876, Rho CrB, and 47 UMa are in [8] and [9].

## 4. CONCLUSION: EXTENSION TO OTHER SYSTEMS

Comparison of the outcomes of orbital integration with Fig. 1 enable some rules of thumb to be established, as follows.

- If $(3R_H + \Delta r)$ for a giant planet extends across, or nearly across, a HZ, there will be no confined orbits in the HZ.

- If $(3R_H + \Delta r)$ extends close to a HZ, or just penetrates it, then there will be confined orbits in some of the HZ
- If $(3R_H + \Delta r)$ stops well short of a HZ then there will be confined orbits in most or all of the HZ. 'Well short' is not yet quantified.
- There can be exceptions to confinement at certain mean motion resonances. (And presumably at certain secular resonances.)

In order to avoid the lengthy task of applying orbital integration to all the other exoplanetary systems, we will shortly be applying these rules so that a quick assessment will be made of whether an exoplanetary system could have Earth-mass planets in confined orbits (almost) anywhere, somewhere, or everywhere in the HZs. In the meantime, a *crude* estimate of the *proportion* of systems that could have confined orbits somewhere in a HZ can be obtained from Fig. 2.

Fig. 2 shows the (minimum) masses of the 107 confirmed planets around main sequence stars, versus their *normalised* distances from their stars. These distances have been scaled so that HZ(0) for the Sun, shown by the grey strip, can be applied approximately to all of them. For this adjustment, stars have been grouped as follows: 0.32 solar masses (Gl876), 0.4-0.6 solar masses, 0.6-0.8 solar masses, 0.8-1.2 solar masses (no adjustment), 1.2-1.4 solar masses.

Planets shown by black discs ($e < 0.1$) and black ellipses ($e > 0.1$) have not been investigated by long-term integrations. Those represented by asterisks (elongated if $e > 0.1$) have no confined orbits in the HZs. Those represented by open symbols have confined orbits (almost) everywhere in the HZs, and those represented by grey symbols have confined orbits in variously restricted regions in the HZs. The correspondence between these outcomes and the location of HZ(0) is reasonable, bearing in mind the crudeness of the grouping and the neglect of the giant planet masses and the large spread of eccentricities.

On the basis of Fig. 1 we rule out the systems with planets in HZ(0). This eliminates 25 planets, and nearly the same number of systems (of the 107 planets, only 26 are known to be in multiple systems). At the other extreme, the giant planets well interior to HZ(0) allow safe havens in their HZs – there are about 30 of these. Of the remainder, some will be ruled out because the giant has high mass, or high eccentricity, or both. We guesstimate that a large proportion is ruled out, and that overall roughly two-thirds of the 93 main sequence exoplanetary systems currently known might be unable to have Earth-mass planets in stable, confined orbits anywhere in the HZs. Our current work will improve on this crude estimate.

Whether the Earth-mass planets could form in the HZs of the exoplanetary systems is a question that needs further study.

## 5. REFERENCES

1. Kasting, J.F., Whitmire D.P., and Reynolds, R.T., Habitable zones around main sequence stars, *ICARUS*, Vol. 101, 108-128, 1993.

2. Private communication with Ulrich Kolb, u.c.kolb@open.ac.uk

3. Chambers, J.E., A hybrid symplectic integrator that permits close encounters between massive bodies, *MNRAS*, Vol. 305, 793-799, 1999.

4. Schneider, J., http://www.obspm.fr/encycl/catalog.html

5. Gonzalez, G., Spectroscopic analyses of the parent stars of extrasolar planetary system candidates, *A&A*, Vol. 334, 221-238, 1998.

6. Laughlin, G., Chambers, J.E., & Fischer, D.A., A dynamical analysis of the 47 UMa planetary system, *APJ*, Vol. 579, 455-467, 2002.

7. Noble, M., Musielak, Z.E., & Cuntz, M., Orbital stability of terrestrial planets inside the habitable zones of extrasolar planetary systems, *APJ*, Vol. 572, 1024-1030, 2002.

8. Jones, B.W., Sleep, P.N., & Chambers, J.E., The stability of the orbits of terrestrial planets in the habitable zones of known exoplanetary systems, *A&A*, Vol. 366, 254-262, 2001.

9. Jones, B.W., & Sleep, P.N., The stability of the orbits of Earth-mass planets in the habitable zone of 47 Ursae Majoris, *A&A*, Vol. 392, 1015-1026, 2002.

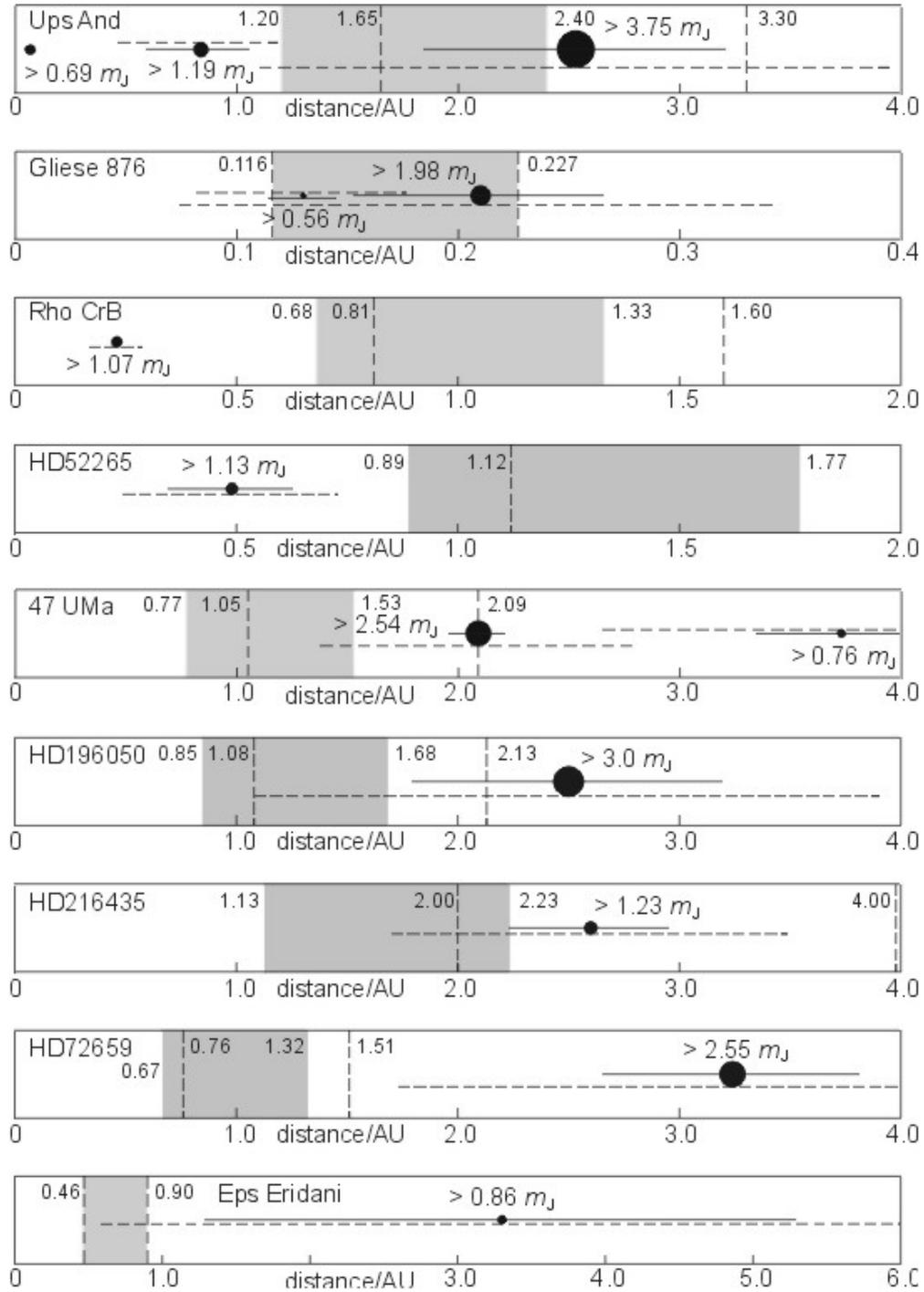

Fig. 1. The nine exoplanetary systems studied, ordered by increasing period of innermost planet. At each giant planet (solid disc) the solid line shows the total excursion $2\Delta r$ of the giant due to its eccentricity. The dashed line extends to $(3R_H + \Delta r)$ each side of the giant when it has its minimum mass. The stars are all main sequence, with masses (in solar masses), [Fe/H], and ages (in Ma) as follows. Ups And 1.3, 0.09, 3300. Gl876 0.32, 0, (unknown). Rho CrB 0.95, -0.19, 6000. HD52265: 1.13, 0.11, 4500. 47 UMa 1.03, -0.08, 7000. HD196050 1.1, 0-0.25, 5000. HD216435 1.25, 0.15, 5000. HD72659 0.95, -0.14, <4500. Eps Eri 0.8, -0.1, 500-1000.

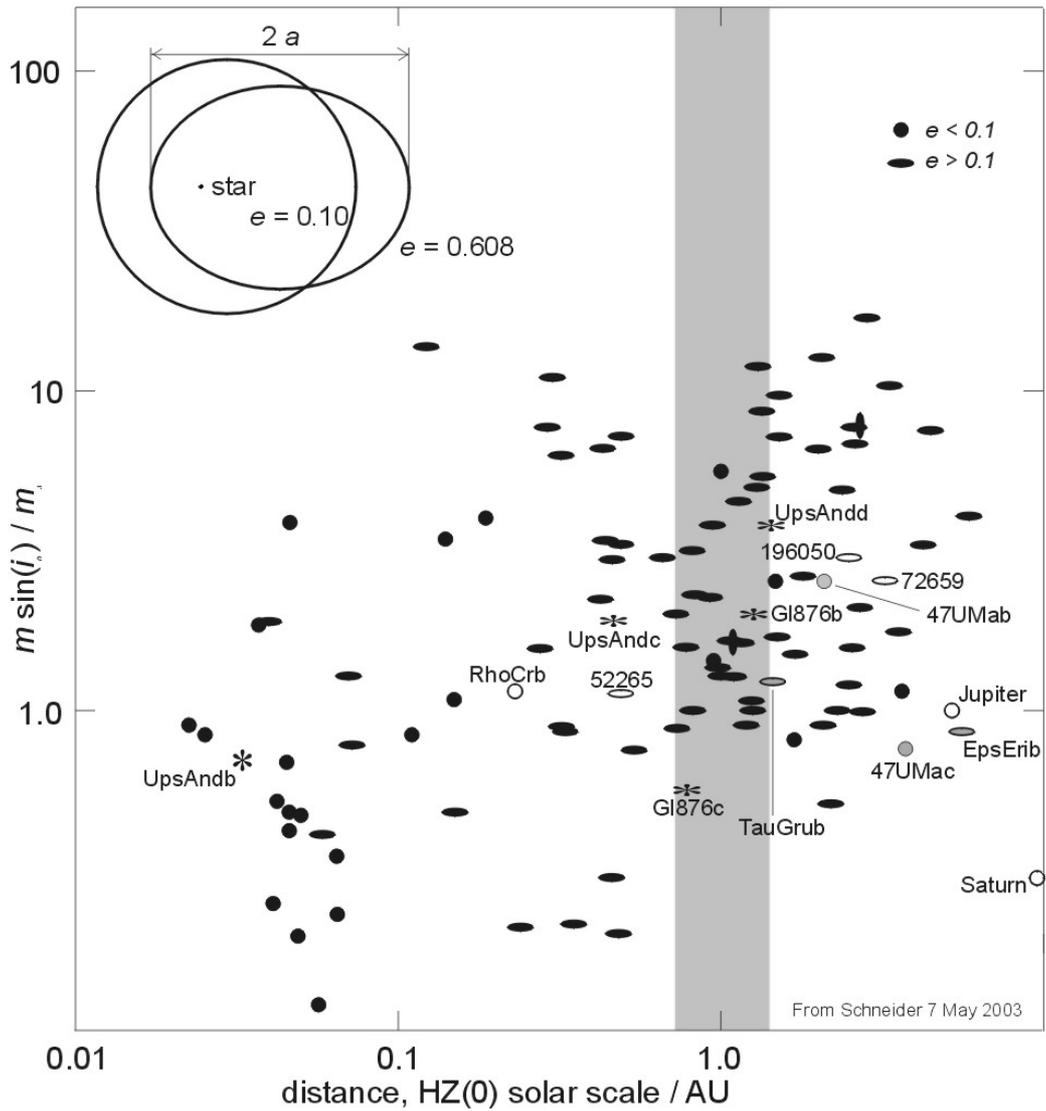

Fig. 2. The (minimum) masses of the 107 confirmed planets around main sequence stars, versus their **_normalised_** distances from their stars. The distances of the planets have been scaled so that HZ(0) for the Sun, shown here by the grey strip, can be applied approximately to all of them. For this adjustment, stars have been binned as follows: 0.32 solar masses (Gl876), 0.4-0.6 solar masses, 0.6-0.8 solar masses, 0.8-1.2 solar masses (no adjustment), 1.2-1.4 solar masses. Planets shown by black discs ($e < 0.1$) and black ellipses ($e > 0.1$) have not been investigated by long-term integrations. Those represented by asterisks (elongated if $e > 0.1$) have no confined orbits in the HZs. Those represented by open symbols have confined orbits (almost) everywhere in the HZs), and those represented by grey symbols have confined orbits in variously restricted regions in the HZs. The orbit shown top left with $e = 0.608$ has the eccentricity of Epsilon Eridani b.